\documentstyle[aps,preprint,tighten,epsf]{revtex}

\begin{document}

\title{`One Sided' Log-normal Distribution of 
Conductances for a Disordered Quantum Wire} 
\date{\today}
\author{K.\  A.\  Muttalib$^{1}$ and 
P.\ W\"olfle$^{2}$}
\address{$^1$Department of Physics, University of 
Florida, P.O. Box 118440, Gainesville,
Florida 32611-8440 \\$^2$Institut f\"ur Theorie 
der Kondensierten Materie, Universit\"at Karlsruhe, 
D-76128 Karlsruhe, and Institut f\"ur Nanotechnologie, 
Forschungszentrum Karlsruhe, Germany.}
\maketitle
\begin{abstract}
We develop a simple systematic method, valid for 
{\it all} strengths of disorder, to obtain 
analytically for the first time the full distribution  of 
conductance $P(g)$ for a quasi one dimensional 
wire in the absence of electron-electron interaction. 
We show that in the crossover region between the 
metallic and insulating regimes, $P(g)$ is highly 
asymmetric, given by an essentially `one sided' 
log-normal distribution. For larger disorder, the tail of the 
log-normal distribution for $g > 1$ is cut off 
by a Gaussian.
\end{abstract}
\draft

Since the discovery of the absence of self averaging 
in mesoscopic disordered systems \cite{Altshuler}, 
the study of the full distribution of conductance 
has attracted a lot of attention \cite{Altshuler1}. 
In particular, while the metallic regime is well 
described by a Gaussian distribution, the moments of 
the conductance fluctuations become of the same order
of magnitude 
as the average conductance on approaching the 
localized regime. In such cases the average value  
becomes insufficient in describing properties of
disordered conductors and the full distribution must 
be considered. 
Recently, numerical support for the existence of a new 
{\it universal} distribution at the metal-insulator 
transition \cite{Shlovskii}, a broad 
distribution of the critical conductance at the 
integer
quantum Hall transition \cite{Wang}, as well as the 
expected multifractal properties associated with 
the critical regime \cite{Wegner} have increased the 
interest in the conductance distribution 
in the intermediate regime, between the well studied 
universal conductance fluctuations in the metallic 
limit and the log-normal distribution in the deeply 
insulating limit.  However,  even for a
quasi one dimensional (1d) system where there is 
only a smooth crossover between the metallic and 
insulating regimes, there is no analytic result 
available for the  conductance distribution in the 
crossover regime. So far only the 
first two moments have been obtained for all strengths 
of disorder \cite{Zirnbauer}, using the 1d 
supersymmetric nonlinear $\sigma$ model \cite{Efetov}. 
This model has been shown \cite{Frahm} to be 
equivalent, in the thick wire or quasi 1d limit, to 
the Dorokov-Mello-Pereyra-Kumar (DMPK) equation 
\cite{Dorokov} which describes the evolution  
of the  distribution of the transmission eigenvalues 
with increasing wire length.

In this work we develop a simple systematic method to 
evaluate directly the full distribution of conductance 
for a thick quasi 1d wire (mean free path $l \gg$ the 
width), starting from the solution 
of the DMPK equation. The main result of the paper is 
that although there is no phase transition in quasi 
one dimension, the crossover region between metallic 
and insulating regimes is highly non trivial, and 
shows a remarkable `one-sided' log-normal 
distribution. Recent numerical studies of a quasi 1d 
system in the quantum Hall regime have shown 
highly asymmetric log-normal distributions in the 
crossover region\cite{Plerou}. We expect similar 
qualitative features to exist in the 
critical regimes in higher dimensions 
as well. Indeed, numerical studies near the integer 
quantum Hall transition in two dimensions as well as 
the Anderson transition in three dimensions also point to 
asymmetric distributions of the critical 
conductance \cite{Ohtsuki}. In addition, we 
predict that even the insulating regime should have a 
sharp cutoff in its log-normal tail near the 
(dimensionless) conductance $g\sim 1$. In particular, 
we show that the conductance distribution in the 
insulating regime 
(in the absence of time reversal symmetry) has the 
form
\begin{equation} 
P[\ln(g)]\approx\cases {\sqrt{\frac{x_1\sinh 2x_1}
{1-g}} e^{-\Gamma x_1^2},& $g < 1$;\cr 
\sqrt{2}ge^{-a(g-1)^2}, & $g\ge 1$ \cr}
\end{equation}
where $x_1=\cosh^{-1}(1/\sqrt{g})$ and the 
parameter $\Gamma=\xi/L$, where $\xi=Nl$ is the 
quasi 1d localization length, $N$ is the number of 
transmission channels, and $L \gg l$ is the length of 
the conductor. The parameter $a$ is the value of  
$F^{\prime\prime}$ given in (10) evaluated at 
$x_2=2/\pi\Gamma$, and tends to 
$\frac{3}{8}\exp [8/\pi \Gamma]$ for $\Gamma \ll 1$
in insulators.  Note that for 
$g \ll 1$, $x_1\sim \ln(2/\sqrt{g})$, and the 
distribution is log-normal, centered at $-\ln g=1/\Gamma$. 
However, for $g > 1$, 
the tail is cut off by an exponential function over 
an exponentially narrow scale in  $1/\Gamma$. The 
results for 
two different values of $\Gamma$, $0.7$ and 
$0.2$, are plotted in 
fig. 1. The main point is that for very 
strong disorder, the typical value of $g$ is much 
smaller than unity, so the peak of the distribution 
$P[\ln(g)]$ is very far away from $g\sim 1$. In this 
case the exponential cutoff at $g\sim 1$ is less 
relevant. However, even for large disorder, a sharp 
cutoff in the tail for $g > 1$  always exists, as 
shown for $\Gamma=0.2$ in fig. 1. 
At intermediate strength of 
disorder, the 
peak of the distribution is close to the cutoff, and 
the distribution becomes highly asymmetric. In 
particular near the crossover between metallic and 
insulating behavior, the peak is at $g\sim 1$, and 
we obtain a `one-sided' log-normal distribution, as 
shown for $\Gamma=0.7$ in figure 1.

As a check of the scope and validity of the method 
developed here, we obtain the exact universal 
conductance fluctuation in the weakly disordered 
metallic regime  with the expected Gaussian 
distribution  as well as the 
correct mean and variance of the log-normal 
distribution in the strongly disordered 
localized regime \cite{Altshuler1} within the same 
unified framework. Systematic 
corrections as a function of disorder can be 
obtained from both the metallic and the insulating 
limits. Note however that the analytic results 
presented here near the crossover regime are only 
semi-quantitative, due to the approximate 
analytical evaluations of certain integrals. A more 
quantitative result is possible based on numerical 
evaluations of these integrals.

We first briefly outline the method. The probability 
distribution $p(\lambda)$ of the $N$ variables 
$\lambda_i$, related to the transmission eigenvalues 
$T_i$ of an N-channel 
quasi 1d wire by $\lambda_i=(1-T_i)/T_i$,  satisfy the 
well known DMPK equation \cite{Dorokov}. The solution of 
this equation can be written in the general form 
\cite{Beenakker1}  
$p({\lambda})\propto \exp[-\beta H({\lambda})]$,
where $H(\lambda)$ may be interpreted as the 
Hamiltonian function of $N$ classical charges at 
positions $\lambda_i$.
The symmetry parameter 
$\beta$=1,2 or 4 depending on the symmetry of the 
ensemble \cite{Altshuler1}. The Hamiltonian depends 
on the parameters $L$, $N$ and $l$ only in the 
combination $\Gamma=Nl/L$. We will consider 
the quasi 1d limit where both $N$ and $L$ approach 
infinity keeping $\Gamma$ fixed. 
The dimensionless conductance in terms of 
$\lambda_i$ is given by
$g=\sum_{i}^{N}\frac{1}{1+\lambda_i}$ \cite{Landauer}.
The distributon of conductance can therefore be 
written as
\begin{equation}
P(g)=\frac{1}{Z}\int_{-\infty}^{\infty}
\frac{d\tau}{2\pi}\int_{0}^{\infty}\prod_{i=1}^{N}
d\lambda_i  \exp\left[i\tau (g-\sum_{i}^{N}\frac{1}
{1+\lambda_i})-\beta H\right],
\end{equation} 
where $Z$ is a normalization factor. In the metallic 
regime $g \gg 1$, the $\lambda_i$ 
are all very close to each other so that a continuum 
description can be used with a density of $\lambda$ 
finite between zero and an upper cutoff given by the 
normalization condition. This approximation describes 
the universal conductance fluctuations in the metallic 
regime  \cite{Muttalib}. In the deeply insulating 
regime on the other hand, all $\lambda_i$ are 
exponentially large and separated exponentially from 
each other, and the conductance is dominated by the 
smallest eigenvalue. This approximation describes 
the log-normal distribution in the deeply insulating 
regime \cite{Pichard}. It is clear however that none 
of the above descriptions can be used in the crossover 
regime, where the smallest eigenvalue is neither zero, 
nor exponentially large. Nevertheless, it turns out 
that it is possible to  combine the essential features 
of the two descriptions and develop a simple and 
systematic procedure to study the conductance 
distribution at intermediate regimes. For simplicity, 
we will discuss the case $\beta=2$ only.

The basic idea is the following: 

1) We first separate out the lowest eigenvalue 
$\lambda_1$ and treat the rest as a continuum with a 
lower bound at $\lambda_2 > \lambda_1$. Note that this 
approximation can be systematically improved by 
separating out the lowest $n > 1$ eigenvalues and 
treating the rest as a continuum. 

2) The continuum part can be written as a functional 
integration on the generalized density $\rho(\lambda)$, 
and the distribution (2) can be rewritten as
\begin{equation}
P(g)=\frac{1}{Z}\int_{-\infty}^{\infty}
\frac{d\tau}{2\pi}e^{i\tau g}\int_{0}^{\infty}d\lambda_1
\int_{\lambda_1}^{\infty}d\lambda_2
\int D[\rho(\lambda)]
\exp[-F(\lambda_1,\lambda_2;\rho(\lambda);\tau)].
\end{equation}
Here the `Free energy'  
\begin{equation}
F(\lambda_1, \lambda_2; \rho(\lambda);\tau)
=\beta H(\lambda_1, \lambda_2; {\rho(\lambda)})
+ i\tau\left[\frac{1}{1+\lambda_1}
+ \int_{\lambda_2}^{b}d\lambda 
\frac{\rho(\lambda)}{1+\lambda}\right]
\end{equation}
contains the `edge' separating out $\lambda_1$  as well 
as the  `source' terms proportional to $\tau$, plus 
the continuum version of the Hamiltonian of the form
$H=\sum_{i<j}^N u(\lambda_i,\lambda_j)+\sum_i^N V(\lambda_i)$.
The upper limit $b$ is given by the number 
conservation $\int_{\lambda_2}^{b}d\lambda \rho(\lambda)=N-1$.

3) We obtain the density by minimizing the Free energy with 
respect to $\rho(\lambda)$, keeping $\lambda_1$ and 
$\lambda_2$ fixed. This gives
\begin{equation}
\int_{0}^{\infty}d\lambda^{\prime} u(\lambda +\lambda_2,
\lambda^{\prime}+\lambda_2)
\rho_{sp}(\lambda +\lambda_2)=2 V_{tot}(\lambda +\lambda_2),
\end{equation}
where we have shifted the lower limit to zero, and
$V_{tot}(\lambda)=V(\lambda)+\frac{i\tau/\beta}{1+\lambda}
+u(\lambda_1,\lambda)$.
After taking a derivative on both sides 
of (5), the kernel can be 
inverted to obtain the saddle point density 
$\rho_{sp}(\lambda)$. 

4) From the density, we obtain the saddle point Free 
energy
\begin{equation}
F_{sp}= \frac{\beta}{2}\int_{\lambda_2}^{b}d\lambda 
V_{tot}(\lambda)\rho_{sp}(\lambda)+\beta V(\lambda_1)
+\frac{i\tau}{1+\lambda_1}.
\end{equation}

5) Since $V_{tot}$ and therefore $\rho_{sp}$ are 
both linear in $\tau$, the Free energy is quadratic in 
$\tau$ and can be written in the form 
$F_{sp}=F^0+(i\tau)F^{\prime}+\frac{(i\tau)^2}{2}
F^{\prime\prime}$.
The integral over $\tau$ in eq. (3) can then be done 
exactly. The result is
\begin{equation}
P(g)=\frac{1}{Z}\int_{0}^{\infty}d\lambda_1
\int_{\lambda_1}^{\infty}d\lambda_2 e^{-S};\;\;\;
S=-\frac{(g-F^{\prime})^2}{2F^{\prime\prime}}+F^0.
\end{equation}

6) At this point, the integrals over $\lambda_1$ and 
$\lambda_2$ can be evaluated numerically. Instead, 
we use saddle point approximation to do the 
integrals in order to obtain an analytic expression 
for $P(g)$. Solving for 
$\frac{\partial S}{\partial \lambda_i}=0$ for $i=1,2$ 
to determine the saddlepoint values of $\lambda_1$ and 
$\lambda_2$, we obtain the distribution as a 
function of  the 
conductance $g$, in terms of the parameter $\Gamma$. 

In the above approach, if we set both $\lambda_1$ 
and $\lambda_2$ equal to zero, we obtain the correct 
universal value $2/15\beta$ for the variance of $g$. 
This is consistent with the picture that in the 
metallic regime, the eigenvalue density can be treated 
as a continuum from zero to an upper cut off $b$. As 
disorder is increased beyond the metallic regime, this 
picture starts to break down. In particular, the 
smallest eigenvalue is pushed to a finite distance 
from zero depending on the strength $\Gamma$, so 
that the continuum picture 
at the edge no longer holds. The correction to the 
metallic behavior is captured in the present 
approach by evaluating the shifts in $\lambda_1$ and 
$\lambda_2$ within a variational scheme. Since the 
insulating regime is dominated by the smallest 
eigenvalue, this approach clearly captures the correct 
insulating behavior. In the crossover regime, both the 
separation of the smallest eigenvalue as well as the 
rest of the continuum become important. Note that if 
more accuracy is needed, one can in principle separate 
out more than one eigenvalue. 

We now give some details. From the exact solution of 
DMPK eqn., the two and one body terms in the 
Hamiltonian of (2) are known to be \cite{Beenakker1}
\begin{equation}
u(\lambda,\lambda^{\prime})=
-\frac{1}{2}\ln\vert (\lambda-\lambda^{\prime})
(x^2(\lambda)-x^2(\lambda^{\prime}))\vert;
\;\;\; V(\lambda)=\frac{\Gamma}{2}x^2(\lambda),
\end{equation}
where $x(\lambda)=\sinh^{-1}\sqrt{\lambda}$.
Note that the difference $\Delta u=u(\lambda+\lambda_2,
\lambda^{\prime}+\lambda_2)- u(\lambda ,\lambda^{\prime})$ 
is negligible in the insulating 
regime and is a small correction proportional to 
$\lambda_2$ in the metallic regime. Therefore to a 
first approximation, the shifted kernel 
$u(\lambda+\lambda_2,\lambda^{\prime}+\lambda_2)$ can be 
replaced by the 
unshifted kernel  $u(\lambda,\lambda^{\prime})$. One can 
then use the 
saddle point density obtained from the unshifted 
kernel to calculate the correction due to the change 
in the kernel from the shift. This can be rewritten as an 
additional term $\lambda_2 V_2+V_{tot}=V_{eff}$ with an 
unshifted kernel in eq. (5). The unshifted kernel can then 
be inverted to give the saddle point density 
\begin{equation}
\rho_{sp}(\lambda+\lambda_2)=\frac{1}{\lambda(1+\lambda)}
\int_{-\infty}^{\infty}d\lambda^{\prime} K^{-1}(x(\lambda)
-x(\lambda^{\prime}))\frac{d}{d\lambda^{\prime}}V_{eff}
(\vert \lambda^{\prime} \vert +\lambda_2),
\end{equation}
where the inverse of the unshifted kernel is 
$K^{-1}(t)=-(1/2\pi^2)\int_0^{\infty}dq \sin (qt) 
(1-e^{-\pi q})$.
The condition $\rho_{sp}(\lambda) \ge 0$ for all $\lambda$ 
in eq. (9) requires  
$\lambda_2- \lambda_1 >\lambda_c=(2/\Gamma\pi)^2$.
The free energy can now be obtained from eq. 
(6).  The integrals for $F^{\prime\prime}$ can be 
done exactly. In terms of the  variables $x_1$, $x_2$, 
defined as 
$\sinh^2 x_i=\lambda_i, i=1,2$, we get
\begin{equation}
F^{\prime\prime}(x_2)= \frac{1}{\sinh^2 2x_2}
\left[-\frac{1}{3}+\frac{1}{4x_2^2}
-\frac{1}{\sinh^2 2x_2}\right].
\end{equation} 
The integrals for $F^{\prime}$ and $F^0$ can be 
evaluated analytically in two limits. For 
$x_2 \ll 1$,
$$
F^{\prime}\approx \Gamma -b_1 \Gamma x_2^2
+\frac{32}{\pi^3}\sqrt{x_2^2-x_1^2};
$$
\begin{equation}
F^0\approx \frac{3\pi^2}{8}\Gamma^2 x_2^2
-2\pi \Gamma \sqrt{x_2^2-x_1^2}
+\frac{3}{2}\ln(x_2^2-x_1^2)-\ln x_1,
\end{equation}
where $b_1\approx 0.89$. In the other limit 
$x_2 \gg 1$,
\begin{equation}
F^{\prime}\approx\frac{1}{\cosh^2 x_1}; \;\;\;
F^0\approx\Gamma x_1^2
-\frac{1}{2}\ln(x_1 \sinh 2x_1)+
\frac{1}{3}\Gamma^2 x_2^3-\Gamma x_2^2+x_2.
\end{equation}
In the metallic regime, $\Gamma \gg 1$, and $x_2$ 
can be very small. Then eqs (10) and (11) give the 
correct mean conductance $\left<g\right>=\Gamma$ and 
variance $var(g)=1/15$. When 
$\Gamma < 1$, $\lambda_2-\lambda_1 > \lambda_c$
requires $x_2 \gg 1$.  In this case the limit 
$x_1 \gg 1$ corresponds to the insulating limit, but 
the limit $x_1 \ll 1$ corresponds to 
the intermediate case close to the crossover regime. 
We therefore study this regime within a saddle point 
approximation for the integrals (7). 

The condition $\frac{\partial S}{\partial x_1}=0$  
has the solution $\cosh x_1^{sp}
=\frac{1}{\sqrt{g}}$, while the condition 
$\frac{\partial S}{\partial x_2}=0$ has the solution 
$x_2^{sp} = 1/\Gamma$. This leads to the saddle 
point result $S^{sp}$,
to which the contributions from the fluctuations 
$S^{fl}=\ln \vert \partial^2 S/\partial x_1^2\vert$
have to be added, leading to eq. (1). In the deeply 
insulating regime $\Gamma \ll 1$, the 
above expression leads to the known mean and variance  
$\left<\ln(1/g)\right>=var[\ln(g)]/2=1/\Gamma$.
However, note that since $\cosh x_1 \ge 1$, the 
saddle point solution exists only for $g \le 1$. For 
$g > 1$, the $x_1$ and $x_2$ integrals are 
dominated by the boundary values at $x_1=0$ and 
$x_2=2/\pi\Gamma$, which has been incorporated in (1). 
In fig. 2 we show $P(g)$ as obtained from eq. (1) for 
$\Gamma=0.7$. The skewed shape and 
the exponential drop at $g \sim 1$ compare well with 
numerical results of \cite{Plerou} for a slightly 
smaller value of $\Gamma=0.5$. The difference in the 
$\Gamma$ values simulates to some extent the 
correction terms to eq. (1) expected for values of 
$\Gamma$ approaching unity. Also shown in fig. 2 is 
the result of a numerical integration of eq. (7) 
using eq. (11) for $\Gamma=1.6$ in the 
metallic regime. The Gaussian shape of $P(g)$ obtained 
for this rather small value of $\Gamma$ is in very 
good agreement with the results of \cite{Plerou}.

We note that according to the relation 
$g=\sum \frac{1}{1+\lambda_i}$, any non-negligible 
$P(g\sim 1)$ comes from the possibility that the 
smallest eigenvalue $\lambda_1$ can be close to the 
origin. However, given $\lambda_1 \ll 1$, the 
logarithmic repulsion between eigenvalues generated 
from (8) forces the rest of the eigenvalues 
exponentially far when $\Gamma \ll 1$, so $P(g > 1)$ 
is cutoff sharply. Since these arguments are quite 
general, we expect qualitatively similar features in 
higher dimensions as well, which should have important 
consequences for the universal conductance 
distribution in the critical regime.

To summarize, we calculated the distribution of 
conductances $P(g)$ for a quasi 1d disordered system
using known results for the DMPK equation. In this
case $P(g)$ depends only on one
parameter $\Gamma=\xi/L$, where $\xi$ is the 
localization length. In the crossover regime 
$\xi/L \sim 1$, we find that $P(g)$ is given by a 
`one-sided' log-normal distribution, cut off by a 
Gaussian tail on the metallic side ($g > 1$). We
believe that this behavior could be generic for  
$P(g)$ in the transition regime even in higher 
dimensions, provided the average of $g$ at the
transition or crossover region is of order unity.
Our results can not be directly compared
to the work of \cite{Altshuler2} in $d=2+\epsilon$
dimensions, because for $\epsilon \ll 1$, 
$\langle g \rangle = 1/\epsilon$ is large, and the 
bulk of $P(g)$ is located deep in the metallic regime.
As proposed in \cite{Shapiro}, the center of $P(g)$
is then Gaussian, with power law tails 
$\propto g^{-2/\epsilon}$. The latter results are 
peculiar to the behavior in $2+\epsilon$ dimensions.
One should keep in mind that the DMPK approach does not 
contain the effects of wave function correlations in the 
transverse direction, which are expected to be important
in higher dimensions. Nonetheless, the similarities of the
shape of $P(g)$ in the crossover regime obtained here with 
the numerically determined $P(g)$ in 3d at the 
critical point \cite{Ohtsuki} appears to suggest that the 
generic behavior 
of $P(g)$ is that of a log-normal distribution for $g < 1$ 
combined with a Gaussian cut off for $g > 1$.

We are grateful to A. Mirlin for stimulating 
discussions and bringing refs. \cite{Plerou} and 
\cite{Ohtsuki} to our attention. We also thank M. 
Fogelstroem for his help regarding numerical 
evaluations. This work has been 
supported in part by SFB 195 der Deutschen 
Forschungsgemeinschaft.

\newpage

FIGURE CAPTIONS:

Figure 1: Log-normal distribution of conductance 
$P[ln(g)]$   
given by eq. (1) in the insulating regime for two 
strengths of disorder, $\Gamma =0.2$ (dashed line) 
and $\Gamma =0.7$ (solid line). 

Figure 2: Distribution of conductance $P(g)$ on the 
insulating and metallic sides of the crossover 
regime ($\Gamma \sim 1$) for two strengths of 
disorder, $\Gamma =0.7$ (solid line) and 
$\Gamma =1.6$ (dashed line).


\begin{references}
\bibitem{Altshuler}
B.L. Altshuler, JETP Lett. 41, 648 (1985); P.A. Lee 
and A.D. Stone, Phys. Rev. Lett. 55, 1622 (1985). 
\bibitem{Altshuler1} 
For reviews, see {\em Mesoscopic 
Phenomena in Solids} 
ed. B.L. Altshuler, P.A. Lee and R.A. Webb, Elsevier, 
Amsterdam 1991; 
C.W.J. Beenakker, Rev. Mod. Phys. 69, 731 (1997).
\bibitem{Shlovskii}
B.~I. Shklovskii et al, Phys. Rev. B47, 11487 
(1993); P. Markos and B. Kramer, Phil. Mag. B68, 357 
(1993).
\bibitem{Wang}
Z. Wang, B. Jovanovic and D-H. Lee, 
Phys. Rev. Lett. 77, 4426 (1996); S. Cho and M.P.A. 
Fisher, Phys. Rev. B55, 1637 (1997); D.H. Cobden and E. 
Kogan, Phys. Rev. B54, R17316 (1997).
\bibitem{Wegner}
For a review, see M. Janssen, Phys. Rep. 295, 1 
(1998).
\bibitem{Zirnbauer}
M. R. Zirnbauer, Phys. Rev. Lett. 69, 1584 (1992);
A. D. Mirlin, A. M\"uller-Groeling and M. Zirnbauer, 
Ann. Phys. (N.Y) 236, 325 (1994).
\bibitem{Efetov}K. B. Efetov and A. I. Larkin, 
Sov. Phys. JETP 58, 444 (1983).
\bibitem{Frahm} 
K. Frahm, Phys. Rev. Lett. 74, 4706 (1995); P. Brouwer 
and K. Frahm, Phys. Rev. B53, 1490 (1996); B. Rejaei,
Phys. Rev. B53, 13235 (1996).
\bibitem{Dorokov}
O. N. Dorokhov, JETP Lett. 36, 318 (1982); P. A. 
Mello, P. Pereyra and N. Kumar, Ann. Phys. (N.Y.) 
181, 290 (1988).
\bibitem{Beenakker1}
C. W. J. Beenakker and B. Rejaei, Phys. Rev. Lett. 71, 
3689 (1993).
\bibitem{Plerou}
V. Plerou and Z. Wang, Phys. Rev. B58, 1967 (1998).
\bibitem{Ohtsuki}
B. Jovanovic and Z. Wang, Phys. Rev. Lett. 81, 2771 
(1998);
K. Slevin and T. Ohtsuki, Phys. Rev. Lett. 78, 4083 
(1997); C. Soukoulis et al, Phys. Rev. Lett. 82, 668 (1999); 
T. Ohtsuki et al, eprint cond-mat/9809221.
\bibitem{Landauer}
R. Landauer, IBM J. Res. Dev. 1, 223 (1957); D. S. Fisher
and P. A. Lee, Phys. Rev. B 23, 6851 (1981); E.N. Economou
and C.M. Soukoulis, Phys. Rev. Lett. 46, 618 (1981).
\bibitem{Muttalib}
K.A. Muttalib, J-L. Pichard, and A.D. Stone, 
Phys. Rev. Lett. 59, 2475 (1987).
\bibitem{Pichard} 
J-L. Pichard et al, J. Phys. France  51, 587 (1990).
\bibitem{Altshuler2}
B. L. Altshuler, V. E. Kravtsov and I. V. Lerner, Sov. Phys. 
JETP 64, 1352 (1986) and Phys. Lett. A134, 488 (1989).
\bibitem{Shapiro}
B. Shapiro, Phys. Rev. Lett. 65, 1510 (1990).
\end{references}
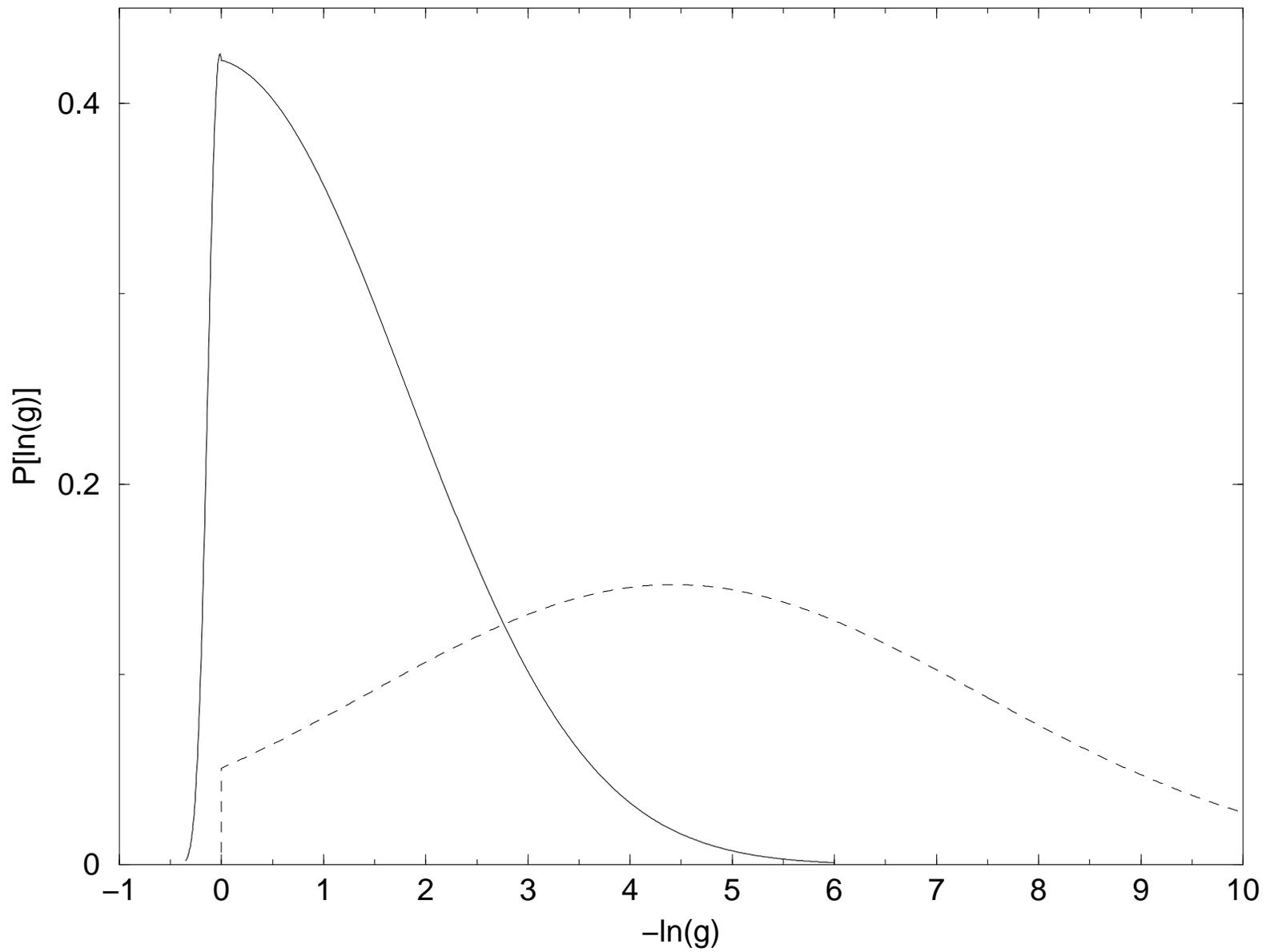
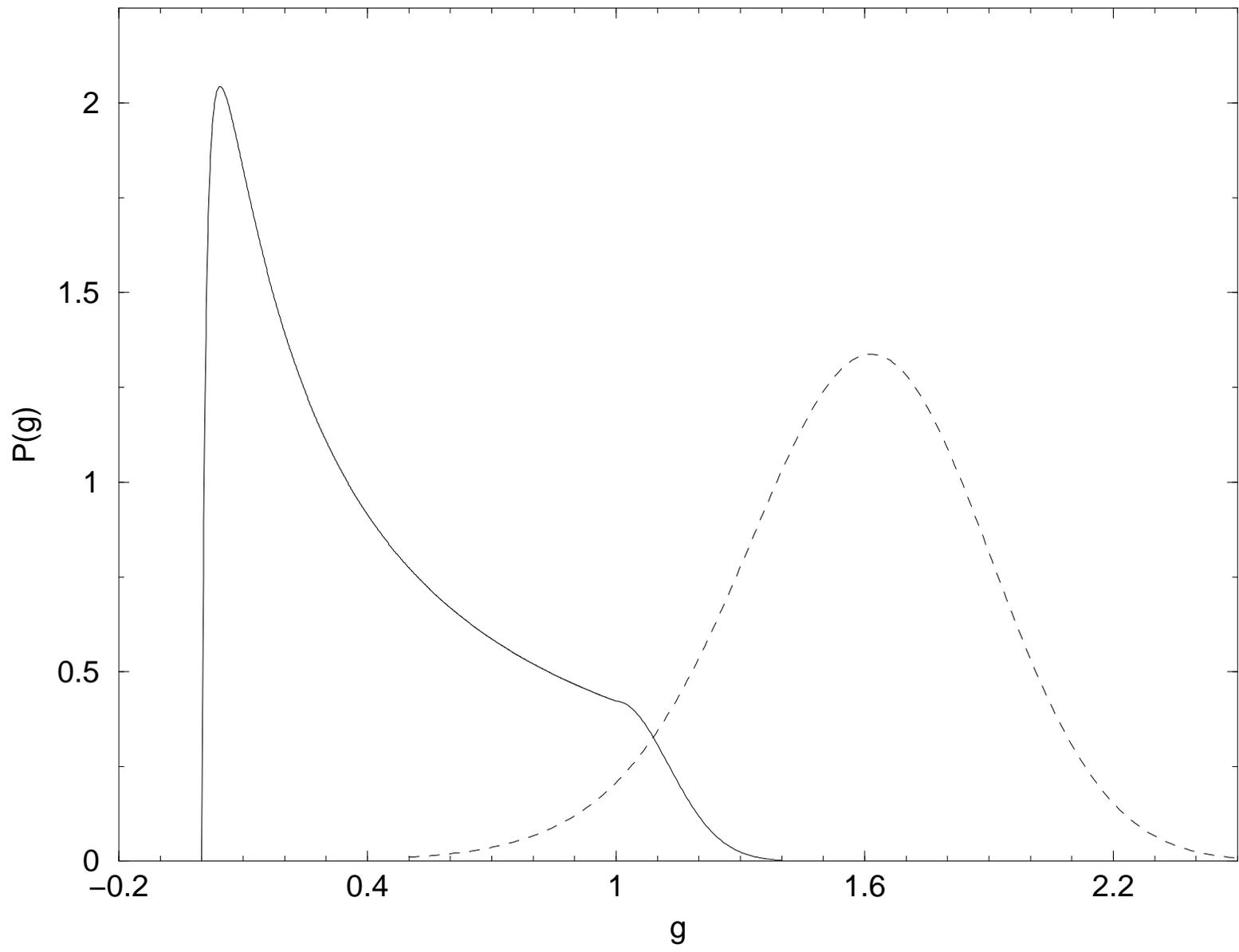
\end{document}